\documentclass[10pt]{article}

\usepackage{amsmath}
\usepackage{amssymb}

\usepackage{graphicx}

\usepackage{cite}

\usepackage{color} 

\usepackage[labelfont=bf,labelsep=period,justification=raggedright]{caption}

\makeatletter
\renewcommand{\@biblabel}[1]{\quad#1.}
\makeatother

\date{}

\begin{document}

% Title must be 150 characters or less
\begin{flushleft}
{\Large
\textbf{Thermoresponsive micropatterned substrates for single cell studies}
}
% Insert Author names, affiliations and corresponding author email.
\\
Kalpana Mandal$^{1}$, 
Martial Balland$^{1,\ast}$, 
Lionel Bureau$^{1,\ast}$
\\
\bf{1} Univ. Grenoble 1/CNRS, LIPhy UMR 5588, BP 87, 38041 Grenoble France.
\\
$\ast$ E-mail: martial.balland@ujf-grenoble.fr, lionel.bureau@ujf-grenoble.fr
\end{flushleft}

% Please keep the abstract between 250 and 300 words
\section*{Abstract}
We describe the design of micropatterned surfaces for single cell studies, based on thermoresponsive polymer brushes. We show that brushes made of poly(N-isopropylacrylamide) grafted at high surface density display excellent protein and cell anti-adhesive properties. Such brushes are readily patterned at the micron scale via deep UV photolithography. A proper choice of the adhesive pattern shapes, combined with the temperature-dependent swelling properties of PNIPAM, allow us to use the polymer brush as a microactuator which induces cell detachment when the temperature is reduced below $32^{\circ}$C.

% Please keep the Author Summary between 150 and 200 words
% Use first person. PLoS ONE authors please skip this step. 
% Author Summary not valid for PLoS ONE submissions.   
%\section*{Author Summary}

\section*{Introduction}

Surface micropatterning is a powerful tool for the design of cell-based assays and sensors, or for fundamental studies of cellular response to environmental cues  \cite{whitesides,review,thery1}. The combination of surface chemistry and microfabrication techniques allows to create substrates onto which adhesion can be tuned so as to obtain regular 2D arrays of immobilized cells. Such patterns have proven to be highly valuable for $e.g.$ statistical analysis of the response of cells cultured in a well-controlled microenvironement \cite{thery2}.

Many different strategies have been developed to fabricate surfaces presenting cell-adhesive patterns, among which the most popular are based on microcontact printing or photolithography \cite{review,thery1}. These widespread techniques may yet exhibit drawbacks in terms of ease of use ($e.g.$ needed equipments or large number of steps), reproducibility, large scale homogeneity of the patterns, or stability of the produced surfaces. A key point in designing such surfaces is to obtain a high contrast between the  regions onto which cells attach and the surrounding non-adhesive background. The use of background polymer coatings, and in particular polymer brushes, have become a favorite choice, for they exhibit excellent protein-repellency, hence efficient cell non-adhesiveness \cite{ma,Huck,racine,thery3}. 

On the one hand, patterned brushes of ``passive''  water soluble polymers have been elaborated via two main routes: 

(i) uniform coating of the substrate by a "grafted-onto" brush, i.e. by adsorption of a block-copolymer containing a protein-repellent part (often poly(ethylene-glycol)) stretching away from the underlying surface. Such a uniform brush is subsequently patterned by selective UV irradiation to create adhesive zones\cite{thery3,racine}. 

(ii) polymer brushes grafted from the substrate, i.e. grown from a layer of polymerization initiators first grafted  on the substrate. Patterning is achieved by micro-contact printing of the initiator, which ensures
a growth of polymer chains restricted to the intiator-printed regions \cite{ma,Huck}. 

These two techniques have already proven to yield patterned substrates suitable, in terms of length scale and adhesive contrast, for single cell studies. However, long term use or storage stability is a limitation of coatings using physisorbed copolymers, while spatial resolution might be an issue with methods employing microcontact printing, because surface diffusion of the printed molecules may blur the initial pattern.

On the other hand, the use of thermosensitive polymer coatings (as opposed to the above ``passive'' ones) attracts an ever-growing interest in the field of cell adhesion control. Since the pioneer work of Okano et al. \cite{okanoroots}, it has been shown that poly(N-isopropylacrylamide) (PNIPAM) brushes could switch from a cell-adhesive to a cell-repellent state when the temperature was decreased from 37$^{\circ}$C to below 32$^{\circ}$C, 
the temperature at which water goes from poor to good solvent of PNIPAM chains \cite{werner,okano}. Such a temperature-controlled cell adhesion has been observed to be favored by brushes of low enough thicknesses, while thicker and/or more dense brushes behaved essentially as non-adhesive coatings irrespective of the temperature \cite{thick1,thick2}. Coatings based on PNIPAM or its copolymers have been employed in several previous studies in order to create patterned thermoresponsive substrates \cite{ito,okano2,okano3,okano4}. However, these studies have focused on patterns of large dimensions (tens or hundreds of microns up to several millimeters), for they aimed at harvesting macroscopic cell sheets for tissue reconstruction applications.

In the present article, we report a method to fabricate thermoresponsive patterned substrates which combines many of the advantages of the above-mentioned techniques and allows for single cell studies.

We show that the use of high density polymer brushes of PNIPAM, bound to glass substrates via the so-called ``grafting-from'' method, and patterned by direct photo-ablation, represents a reliable, fast and cost-effective technique to design thermosensitive micropatterned  platforms.
Compared to the existing well-established templating techniques, the method we report presents the following important features, and comes as an interesting alternative to $e.g.$ elaboration of coatings based on adsorbed ethylene-glycol copolymers \cite{thery3}: 

(i) PNIPAM brushes are elaborated from common and inexpensive chemicals, and their molecular structure can be tuned at will.  Furthermore, micron-scale patterning is achieved in one single step, without requiring access to clean room facilities.  

(ii) Polymer chains being covalently bound to the substrate, such coatings  show excellent usage and storage long-term stability. 

(iii) High grafting density brushes display superior protein and cell repellency, obtained in an extremely reliable and reproducible way.  
 
 (iv) Although such high density brushes are cell-repellent at 37$^{\circ}$C, PNIPAM chains still shift from a collapsed to a swollen state as the temperature is decreased below the polymer LCST (Lower Critical Solution Temperature) of 32$^{\circ}$C. This temperature-controlled conformation change of PNIPAM, combined with a proper choice of the pattern shapes, make the polymer coating act as a thermoactuator which allows us to detach the studied cells by lowering the surface temperature.  This adds a very attractive feature to the usual ``passive'' micropatterned platforms. 

% You may title this section "Methods" or "Models". 
% "Models" is not a valid title for PLoS ONE authors. However, PLoS ONE
% authors may use "Analysis" 
\section*{Materials and Methods}

\begin{figure}[h!]
\centering
  \includegraphics[width=7cm]{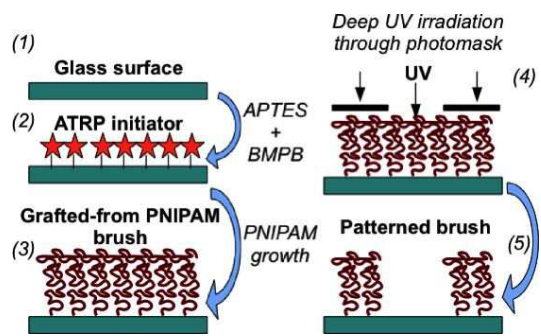}
  \caption{Elaboration steps: grafting of an ATRP initiator on a glass surface (2) is followed by NIPAM polymerization (3), yielding a polymer brush which is selectively removed by UV irradiation (4).}
  \label{fig:principle}
\end{figure}

The various steps required to produce micropatterned PNIPAM brushes are summarized on the scheme Fig. \ref{fig:principle}.

\subsection*{Polymer brush synthesis}

PNIPAM brushes were grafted from glass coverslips and oxidized silicon wafers by surface-initiated Atom Transfer Radical Polymerization (ATRP), according to a protocole akin to that described in details in \cite{bibi}. 

N-isopropylacrylamide (NIPAM) was purified by recrystallization in n-hexane. 
3-aminopropyl-triethoxysilane (APTES),  triethylamine (TEA), copper chloride (CuCl),  1,1,7,7-Pentamethyldiethylenetriamine (PMDETA) and 
2-bromo-2-methylpropionyl bromide (BMPB) were used as received. All aqueous solutions were prepared in ultra-pure water. 
(i) Glass and silicon substrates were cleaned  in a 1 M sodium hydroxide aqueous solution for 15 minutes and rinsed with water.
(ii) Samples were immersed, for 1 minute, in an aqueous solution of APTES of concentration $c_{\text{APTES}}$ chosen in the range 10$^{-5}$--$2\times 10^{-3}$ M. After rinsing with water and drying in a nitrogen stream samples were immersed, for 1 minute,  in a solution of dichloromethane (25 mL) containing TEA (1.2 mL) and BMPB (260 $\mu$L), followed by rinsing with dichloromethane, ethanol and water. This leads to surface immobilization, on the amino-terminated sites, of the ATRP initiator.
(iii) A solution of  NIPAM (1g), PMDETA (150 $\mu$L) and water (20 mL) was prepared in a flask and bubbled with argon gas for 30 minutes before adding CuCl (25 mg). Initiator-grafted samples were immersed in this solution for a prescribed amount of time during which polymerization occurred, and finally rinsed with pure water.

\subsection*{Brush characterization}

Brushes were characterized by ellipsometry and Surface Forces Apparatus (SFA) mesurements.

We have used a custom-built ellipsometer in the rotating compensator configuration, at a wavelength of 632 nm and an angle of incidence of 70$^{\circ}$. The dry thickness of the brushes grown on oxidized silicon wafers was determined assuming a  Si/SiO$_2$/PNIPAM multilayer, with a thickness of 2 nm and a refractive index of 1.46 for silicon oxide, and a refractive index of 1.47 for the PNIPAM layer \cite{Leckband}.

Surface Forces experiments were performed on a home-built instrument, according to a protocol described in details in \cite{bibi}. Briefly, a pair of freshly cleaved mica sheets ($\sim 5\, \mu$m in thickness) were glued onto cylindrical lenses of 1 cm radius of curvature. A PNIPAM brush was grown on one mica sample after plasma activation of its surface. The brush-bearing mica sheet was then mounted into the SFA, facing the bare mica sample, and the gap between the to surfaces was filled with ultra-pure water (see inset of Fig.  \ref{fig:swelling}). The surfaces were then approached at low velocity ($\sim$1 nm.s$^{-1}$), while recording the  force and the distance between the mica substrates by means of multiple beam interferometry, as described in \cite{bibi}. Force/distance curves during quasi-static compression have been measured at temperatures of 25 and 37$^{\circ}$C.

\subsection*{Patterning}

Dry PNIPAM-bearing coverslips were placed in direct contact with a chromium quartz photomask (Toppan Photomasks inc., Texas USA). UV irradiation of the surfaces through the photomask was done in a custom-built device housing a set of 4 low-pressure mercury lamps (Heraeus Noblelight GmbH, NIQ 60/35 XL longlife lamp, $\lambda=$185 and 254 nm, quartz tube, 60 W). Samples were placed at a fixed distance of 9 cm from the UV tubes and irradiated for a prescribed duration between 5 and 10 minutes.  

\subsection*{Protein coating}

PNIPAM treated glass coverslips were first extensively washed with phosphate-buffered saline (PBS), pH 7.4. Before cell seeding, a 100$\mu$L drop\footnote{This volume was used for a 20x20mm glass coverslip and must be adapted to the coverslip size} of protein solution composed of 20 $\mu$g/ml fibronectin (Sigma)/fibrinogen-Alexa fluor 546 nm (Invitrogen) in 10mM Hepes (pH 8.5) was deposited on a flat piece of parafilm. The patterned substrates were then directly placed on top of the protein solution drop and incubated for one hour at room temperature, protected from external light. They were washed twice with PBS.

\subsection*{Cells seeding, fixing and staining}

Mouse Embryonic Fibrobasts (MEF) were maintained at 37$^{\circ}$C in a humidified atmosphere of 5\% CO2 and 95\% air in Dulbecco's modified Eagle medium (DMEM) containing 10\% bovine fetal serum, 0.2\% peni-streptomycin. Cells were deposited on micropatterned surfaces at a density of 50 000 cells/cm$^2$. Micropattern area was adapted to ensure full spreading of MEF cells on each pattern (900 $\mu$m$^2$). 

After 30 minutes non adherent cells localized in between the patterns were removed by gentle flushing with fresh media. After 2 hours of culture, spread cells were either observed  at room temperature during thermodetachment experiments, or fixed in order to preserve their shapes.
For image averaging, cells were fixed in paraformaldehyde 4\% for 30 min at room temperature. After two washes in PBS, fixed cells were permeabilised 10 mins with Triton X-100 0.2\% in PBS. Then cells were stained with FITC conjugated phalloidin at 2 $\mu$M (Sigma Aldrich). After a last wash in PBS, preparations were mounted in fluoroshield mounting medium.

\subsection*{Microscopy}

 Microscopy experiments were performed using a Nikon Ti-E microscope equipped with an incubator maintaining the temperature at 37$^{\circ}$C. Stained cells were imaged with a Nikon 63X oil objective lens (NA 1.4). For the thermoresponsive experiments, time-lapse sequences were acquired while regulating the temperature of the room between 21 and 30$^{\circ}$C. Phase contrast images of the detaching cells were taken using an Olympus CKX41 microscope equipped with a 10X air objective(NA 0.25) and a 12-bit monochrome camera. 

% Results and Discussion can be combined.
\section*{Results and Discussion}

\subsection*{Brush characterization}

\begin{figure}[h!]
\centering
  \includegraphics[width=7cm]{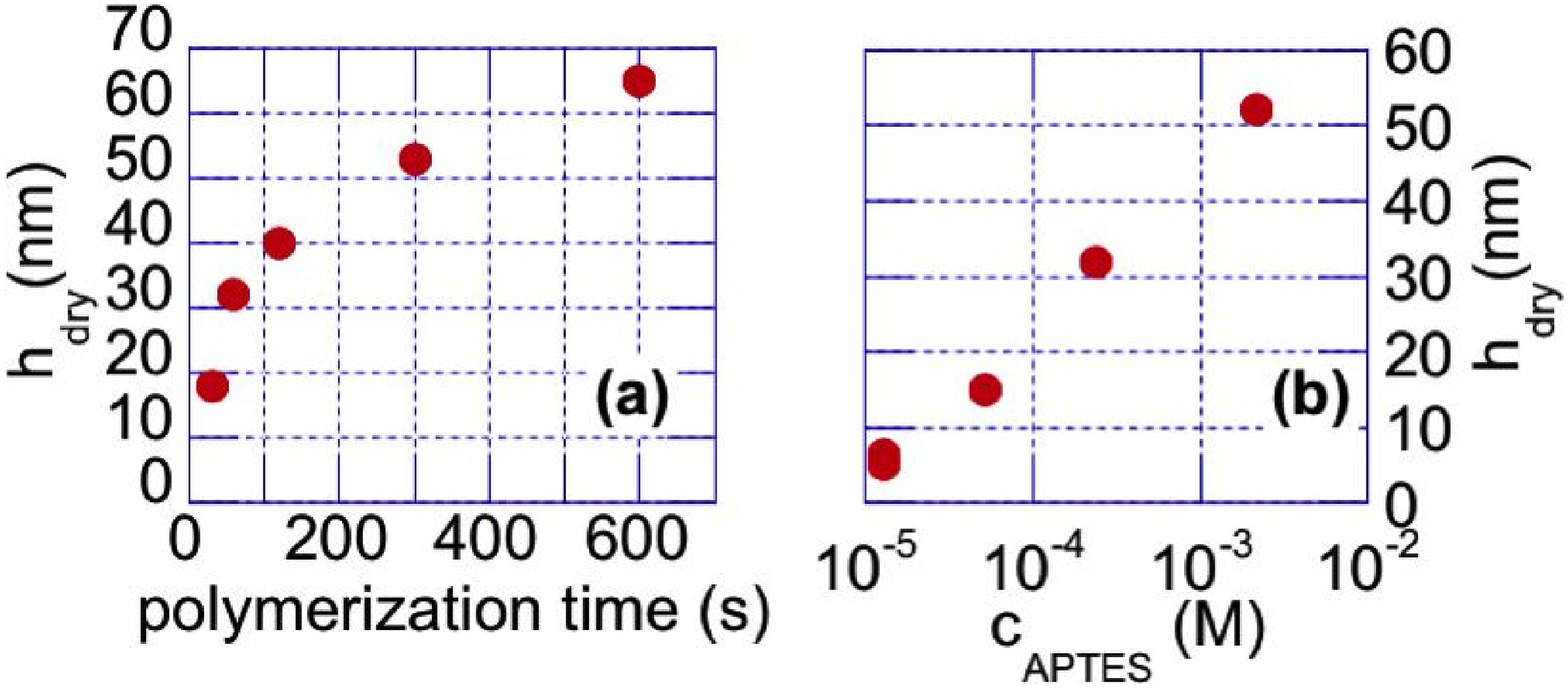}
  \caption{(a) Brush dry thickness $vs$ polymerization time, for  $c_{\text{APTES}}=2.10^{-4}$M. (b) $h_{\text{dry}}$ $vs$ $c_{\text{APTES}}$ for 1 min polymerization.}
  \label{fig:growth}
\end{figure}

PNIPAM brushes grown on silicon wafers were characterized by measuring their dry thickness, $h_{\text{dry}}$ by ellipsometry. The dry thickness of a brush is given by $h_{\text{dry}}=Na^3/d^2$, where $N$ is  the number of monomer per chain, $a$ is the monomer size, and $d$ is the distance between anchoring sites. $N$ is determined by the polymerization time, and $d$ is fixed by the surface density of ATRP initiator, which depends on the concentration $c_{\text{APTES}}$. Fig. \ref{fig:growth}a and b  show that $h_{\text{dry}}$ indeed increases with increasing $c_{\text{APTES}}$ or polymerization time. No difference in  $h_{\text{dry}}$ was noticed between measurements immediately after grafting and after several days of immersion in water, showing that polymer layers are stable and covalently grafted to the underlying substrate. We have checked that all the grafted brushes displayed the previously reported hydrophilic/hydrophobic transition when measuring the water contact angle at temperatures below and above the Lower Critical Solution Temperature (LCST) of PNIPAM. Dry thickness measurements performed at 5-6 different locations over a surface of about 1 cm$^2$ yielded the same results to within $\pm1$ nm, showing that brush growth was very homogeneous over large scales.

\begin{figure}[h!]
\centering
  \includegraphics[width=7cm]{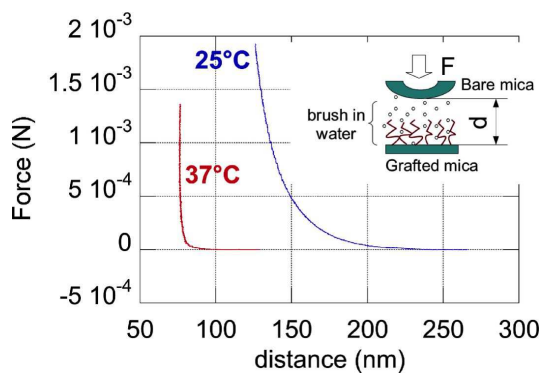}
  \caption{Force-distance curves measured on a water immersed PNIPAM brush ($h_{\text{dry}}=74$ nm), at two temperatures below and above the polymer LCST (see labels on main panel). It can be seen that the range of steric, repulsive forces due to the presence of the brush is markedly reduced at 37$^{\circ}$C, and that the hard-wall repulsion at high temperature occurs at a distance close to the dry thickness of the brush, indicating almost full expulsion of the solvent from the PNIPAM layer above its LCST. Inset: scheme of the SFA experimental configuration.}
  \label{fig:swelling}
\end{figure}

Swelling of the brushes immersed in water was estimated from the Force/Distance curves measured with the SFA.  First, the dry thickness of brushes grown on plasma-activated mica were measured in the SFA  and checked to be the same as those obtained under the same grafting conditions on silicon wafers. This allowed us to control that PNIPAM brushes grown on mica had the same density as those grafted on silicon oxide. Fig. \ref{fig:swelling} illustrates two typical compression curves obtained below and above the LCST: it can be seen that the range of steric repulsive forces is clearly  narrower at 37 than at 25$^{\circ}$C, which indicates chain collapse above the polymer LCST. 

The data shown on Fig. \ref{fig:swelling} have been obtained with a brush of $h_{\text{dry}}=74$ nm, grafted from an initial solution of $c_{\text{APTES}}=2\times 10^{-3}$M. All the results shown in the following have been obtained with brushes of the same grafting density, which we can roughly estimate as follows: as described in Malham and Bureau \cite{bibi}, the grafting density $\sigma$ is expected to be, within the Alexander-de Gennes framework, $\sigma\simeq1/(a \alpha)^2$, with $a$ the monomer size and $\alpha$ the swelling ratio. We compute $\alpha$ as $h_{\text{swell}}/h_{\text{dry}}$, and take for $h_{\text{swell}}$ the distance at the onset of repulsion in SFA experiments performed at room temperature. For the data presented on Fig.\ref{fig:swelling}, $h_{\text{swell}}\simeq 240$ nm at 25$^{\circ}$C, hence $\alpha\simeq 3.2$. Such a swelling ratio is fully consistent with those measured for high density brushes in \cite{bibi}. We then deduce, taking $a=5$\AA, a grafting density $\sigma\simeq 4\times 10^{-3}$ chain.\AA$^{-2}$ for the brushes used in the present study \footnote{This value of $a$ has been estimated from data reported in reference \cite{biggs}. We have used the values of the grafting density, dry thickness and molecular weight reported in \cite{biggs} and computed $a$ from $Na^3=h_{\text{dry}}d^2$}.

\subsection*{Patterns}

\begin{figure}[h!]
\centering
  \includegraphics[width=6cm]{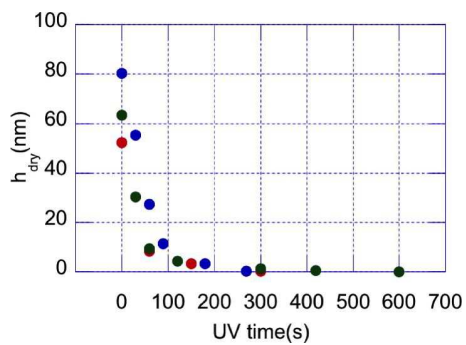}
  \caption{$h_{\text{dry}}$ $vs$ UV irradiation time for brushes of initial thickness 82 nm (blue), 65 nm (green), and 54 nm (red).}
  \label{fig:UVablation}
\end{figure}

Exposure of the dry PNIPAM brushes to deep UV light in air results in ablation of the polymer from the surface\footnote{PNIPAM brushes are stable under soft UV irradiation: we have checked that a 15min exposure at 2 cm from a 75W UV lamp ($\lambda$=365nm) resulted in no dry thickness decrease of the grafted layers.}. We have characterized the ablation rate under such conditions, by monitoring $h_{\text{dry}}$ as a function of UV irradiation time ($t_{\text{UV}}$) on grafted silicon wafers. Results are presented on Fig. \ref{fig:UVablation}. It is seen that, starting from initial brush thicknesses of a few tens of nm, a complete removal of the polymer is achieved for $t_{\text{UV}}\geq 300$s.

\begin{figure}[h!]
\centering
  \includegraphics[width=5cm]{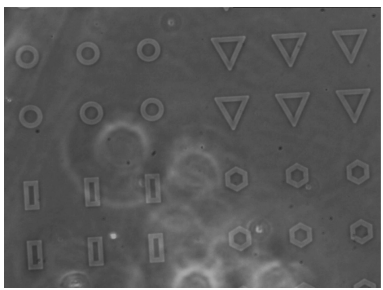}
  \caption{Phase contrast image of annular, triangular, rectangular and hexagonal patterns obtained by UV photoablation of PNIPAM. The light grey regions have been irradiated by deep UV, where the polymer have been removed. Image size is 700$\times$500 $\mu$m$^2$.}
  \label{fig:patternsPhC}
\end{figure}

Patterns elaborated on PNIPAM-bearing coverslips using $t_{\text{UV}}\geq 300$s can be observed by phase contrast microscopy.
Fig. \ref{fig:patternsPhC} provides an illustration of different pattern shapes thus observed. Contrast on such images arises from both the height and the refractive index difference between the PNIPAM background layer and the bare glass which has been exposed in the UV-irradiated regions. Best spatial resolution of the patterns was obtained by placing the dry PNIPAM-bearing coverslips in direct contact with the photomask. This resulted in patterns obtained on PNIPAM being $\simeq 1\, \mu$m broader than the original shapes of the photomask.

Under culture conditions (at 37$^{\circ}$C), the maximum height variation experienced by the cells corresponds to the difference between the bare and polymer-covered regions. This height difference is the collapsed thickness of the PNIPAM brushes (approximately the dry thickness, as suggested by Fig. \ref{fig:swelling}), $i.e.$ at most 70-80 nm in the present study.

\subsection*{Protein and Cell Adhesion}

The results shown below have been obtained with high density brushes ($c_{\text{APTES}}=2.10^{-3}$M) of $h_{\text{dry}}$ varying between 15 and 80 nm, and $t_{\text{UV}}$ between 5 and 10 min. No significant influence of these two parameters on the observed behavior has been noticed. We have used phototomasks displaying various pattern shapes having the same projected area of 900 $\mu$m$^2$. Similar results regarding protein adsorption and cell adhesion have been obtained with freshly elaborated substrates and with samples stored under ambient conditions for three months before use. 

\begin{figure}[h!]
\centering
  \includegraphics[width=6cm]{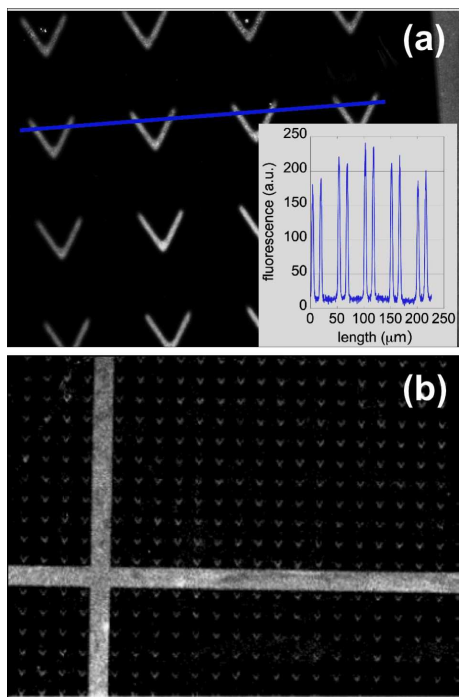}
  \caption{(a) Fibronectin adsorption into V-shaped patterns (V arms of length 40 $\mu$m and width 10 $\mu$m). Inset: fluorescence intensity profile along the blue line drawn in main panel. (b) Wide field image of stained fibronectin adsorbed on V-shaped patterns, showing large scale homogeneity. Image size: 2200$\times$1664 $\mu$m (taken with a 4x objective on an Olympus IX70 microscope. Reduced contrast quality is due to the low NA of the objective)}
  \label{fig:patternProt}
\end{figure}

Images of stained fibronectin adsorbed on the surfaces reveal a high contrast between the UV-irradiated patterns, where the protein adheres, and the brush-covered background which is free of fibronectin. This is illustrated on Fig. \ref{fig:patternProt}: the signal to noise ratio along the intensity profile drawn on Fig. \ref{fig:patternProt}a is about 10:1. This shows that high density PNIPAM brushes are excellent protein-repellent layers at room temperature.

\begin{figure}[h!]
\centering
  \includegraphics[width=8cm]{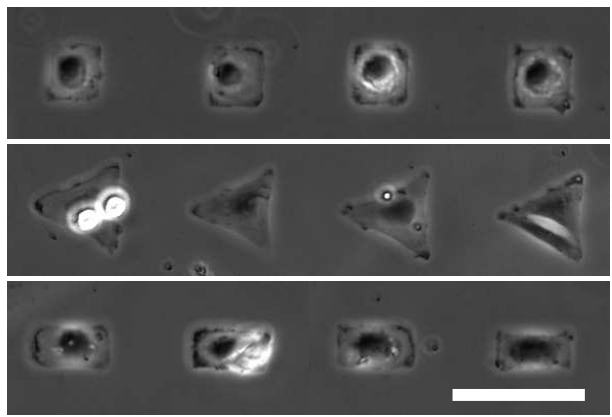}
  \caption{Phase contrast images of cells adhered on square (up), triangular (middle), and rectangular-shaped patterns (down). Scale bar is 80 $\mu$m.}
  \label{fig:patternCell}
\end{figure}

Moreover, the quality of cell patterns obtained at T=37$^{\circ}$C (Fig. \ref{fig:patternCell}) shows that good protein resistance is also maintained above the polymer LCST. Besides, we have checked that, in contrast to the behavior exhibited on dense brushes, cells do adhere, at 37$^{\circ}$C, on low-density brushes ($c_{\text{APES}}=10^{-5}$ M) having the same chain length. Such an effect of brush density on cell adhesion agrees with a recent report \cite{okano}. This indicates that, rather than the absolute thickness of the brush, its grafting density controls protein adsorption and cell adhesion, and not (or to a much lesser extend) the chain length. It is consistent with recent theoretical works concluding that  the protein resistance of brushes is mainly controlled by the osmotic penalty associated with protein insertion within the brush \cite{avi}.

\begin{figure}[h!]
\centering
  \includegraphics[width=8cm]{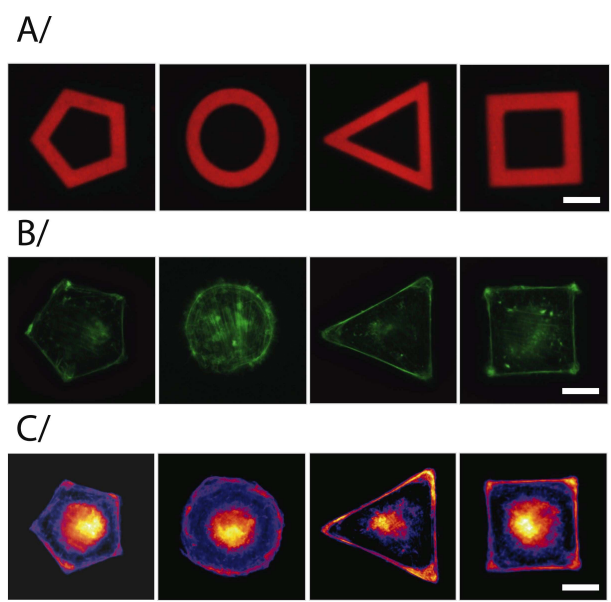}
  \caption{A/ Fibronectin and fibrinogen-A546 coating on micropatterned PNIPAM glass surface (red). Scale bar is 15 $\mu$m.
B/ individual MEF cells plated on pentagon, annulus, triangle or square-shaped fibronectin micropatterns. Cells were fixed and stained with phalloidin to reveal F-actin filaments (green). Scale bar represents 15 $\mu$m.
C/ Average distributions of actin (fire), built from the overlay of 10 images for each shape. The average distribution highlights the reproducibility of the distributions shown in B/ and enhances the spatial distribution of F-actin bundles along micropattern border regions. Scale is 15 $\mu$m}
  \label{fig:actin}
\end{figure}

The photo-ablation technique yields a good pattern resolution: features of typically 10 $\mu$m in size (Fig. \ref{fig:patternProt}a) and down to 5 $\mu$m (Fig. \ref{fig:actin}A) were routinely obtained. This ensures cell shape reproducibility, as shown on Fig. \ref{fig:patternCell}, and makes the present substrates well suited for statistical analysis of cellular response based on image overlay, as exemplified on Fig. \ref{fig:actin}. Previous studies have shown that micropattern allow to control cell cytoskeletal architecture. In particular, cells are known to form actin bundles in response to the geometry of the pattern itself \cite{thery2,Huck}. We have used our PNIPAM patterned substrates to generate averaged actin maps using a home-made software written in Matlab$^{\copyright}$: after normalization of the individual fluorescent images to the same integrated total signal value, averaged fluorescent staining images were automatically aligned, using the protein-stained micropattern images as position references. As can be seen on Fig. \ref{fig:actin}, the heat map generated from the overlay of several actin images unambiguously confirms the ability of our micropatterns to orient actin network organisation: cells form preferentially contractile F-actin bundles, or stress fibers, along the adhesive regions of the micropatterns.

Next, we have checked for the possibility of long term cultures on the substrates. We have maintained cells in culture up to 5 days, during which cell division occurred, indicating good biocompatibility of the patterned surfaces. Furthermore, we have observed that cell adhesion is also achieved without fibronectin pre-coating. Such a non protein-specific cell patterning method, along with the ability to reach long culture time, make the present surfaces a  potentially powerful tool for stem cells culture. It also shows that our technique is a versatile one, for {\it e. g.} different protein coatings can be used on our surfaces, thus allowing to address more specific biological questions.

\subsection*{Thermally induced cell detachment}

\begin{figure}[h!]
\centering
  \includegraphics[width=7cm]{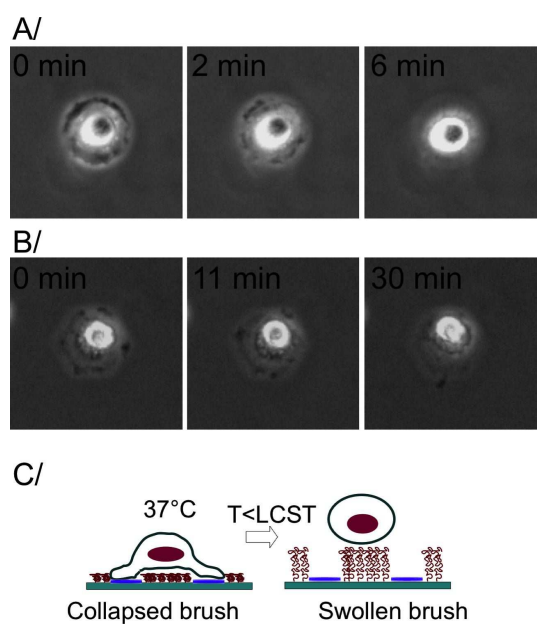}
  \caption{(A) and (B): Sequences of cells detaching as temperature is lowered (images 75$\times$75 $\mu$m, dry brush thickness $h_{\text{dry}}=75$ nm). The time stamp gives the time elapsed since the surfaces were taken out of the incubator. (A) cell initially adhered on a circular pattern. Imposed temperature is 21$^{\circ}$C. (B) cell initially adhered on a hexagonal pattern. Imposed temperature is 26$^{\circ}$C. (C): sketch of the polymer chains swelling inducing cell detachment.}
  \label{fig:detach}
\end{figure}

Finally, we show that, although dense PNIPAM brushes are protein repellent irrespective of the temperature, their thermosensitive property can still be used for local cell manipulation. Fig. \ref{fig:detach} shows a proof of concept for such a thermoresponsive actuation.  It can be seen that cells which were spread and adhered on patterns change their shape to round up and finally detach from the surface within a few minutes after the temperature has been lowered below the LCST. The kinetics depends on the temperature imposed below the LCST, the lower the temperature the faster the detachment. We have checked that such a thermo-actuated detachment does not depend on the pattern shape, provided that the shape is chosen as follows. The pattern has to be such that  cells spread over a non adhesive PNIPAM region, while bridging two adhesion zones, as sketched on Fig. \ref{fig:detach}C. Under such conditions, when the temperature is lowered below the polymer LCST, the grafted chains switch from a collapsed to a swollen conformation, and locally act as a microactuator which generates high enough forces to induce gentle cell detachment, without the need for the trypsin treatment classically employed for cell harvesting. 

Note that the cell detachment mechanism reported here is different from what has been described previously \cite{werner,okano,ito,thick1,thick2,okanoroots}, since it does not involve a temperature-induced change in the cell/PNIPAM affinity, but rather takes advantage of polymer swelling to generate forces.

Cells thus detached from the patterned substrates were subsequently recultured in a polystyrene petri dish and were observed to spread and divide on the surface. This shows that cells detachment achieved by the method we report here does not affect their viability.

\section*{Conclusions}

We have shown that dense PNIPAM brushes exhibit excellent protein resistance and are readily patterned at the micron scale via a single photo-ablation step. The reported fabrication method presents the following advantages: (i) robust and stable anti-adhesive brushes are covalently grafted at high density on common glass coverslips by surface-initiated ATRP, (ii) photolithography yields sharp patterns, in contrast to microcontact-printing techniques which may be limited in resolution by surface diffusion of the printed species \cite{Huck}, (iii) the method is easy to implement and requires only basic laboratory equipment, (iv) patterned substrates can thus be produced within 2-3 hours only, in a highly reproducible way. Moreover, we have shown that a proper choice of the pattern shapes allows us to combine the cell non-adhesiveness of dense PNIPAM brushes with their thermoresponsiveness, which permits gentle cell detachment. These features make such PNIPAM-based substrates a choice tool for single cell patterning and thermally-induced on-chip cell manipulation.

\section*{Aknowledgements}
K.M. is financed by the ``Fondation Nanosciences aux limites de la nanoelectronique'' (contract number FCSN-2008-11JE). We are grateful to P. Ballet for assistance with ellipsometry.

\section*{Author contributions}
M.B. and L.B. conceived and designed the study. K.M., M.B. and L.B. performed the experiments.

%\section*{References}
% The bibtex filename
\bibliography{paperPlos1}

\begin{thebibliography}{10}
\providecommand{\url}[1]{\texttt{#1}}
\providecommand{\urlprefix}{URL }
\expandafter\ifx\csname urlstyle\endcsname\relax
  \providecommand{\doi}[1]{doi:\discretionary{}{}{}#1}\else
  \providecommand{\doi}{doi:\discretionary{}{}{}\begingroup
  \urlstyle{rm}\Url}\fi
\providecommand{\bibAnnoteFile}[1]{%
  \IfFileExists{#1}{\begin{quotation}\noindent\textsc{Key:} #1\\
  \textsc{Annotation:}\ \input{#1}\end{quotation}}{}}
\providecommand{\bibAnnote}[2]{%
  \begin{quotation}\noindent\textsc{Key:} #1\\
  \textsc{Annotation:}\ #2\end{quotation}}
\providecommand{\eprint}[2][]{\url{#2}}

\bibitem{whitesides}
Singhvi R, Kumar A, Lopez GP, Stephanopoulos GN, Wang DI, et~al. (1994)
  Engineering cell shape and function.
\newblock Science 264: 696--698.
\bibAnnoteFile{whitesides}

\bibitem{review}
Falconnet D, Csucs G, Grandin HM, Textor M (2006) Surface engineering
  approaches to micropattern surfaces for cell-based assays.
\newblock Biomaterials 27: 3044--3063.
\bibAnnoteFile{review}

\bibitem{thery1}
Th\'ery M (2010) Micropatterning as a tool to decipher cell morphogenesis and
  functions.
\newblock {J} Cell Sci 123: 4201--4213.
\bibAnnoteFile{thery1}

\bibitem{thery2}
Th\'ery M, Racine V, Piel M, P\'epin A, Dimitrov A, et~al. (2006) Anisotropy of
  cell adhesive microenvironment governs cell internal organization and
  orientation of polarity.
\newblock Proc Natl Acad Sci USA 103: 19771--19776.
\bibAnnoteFile{thery2}

\bibitem{ma}
Ma H, Hyun J, Stiller P, Chilkoti A (2004) `non-fouling'' oligo(ethylene
  glycol)- functionalized polymer brushes synthesized by surface-initiated atom
  transfer radical polymerization.
\newblock Adv Mater 16: 338--341.
\bibAnnoteFile{ma}

\bibitem{Huck}
Gautrot JE, Trappmann B, Oceguera-Yanez F, Connely J, He X, et~al. (2010)
  Exploiting the superior protein resistance of polymer brushes to control
  single cell adhesion and polarization at the micron scale.
\newblock Biomaterials 31: 5030--5041.
\bibAnnoteFile{Huck}

\bibitem{racine}
Racine J, Cheradame H, Chu YS, Thiery JP, Rodriguez I (2010) Micropatterns of
  cell adhesive proteins with poly(ethylene oxide)-block-poly(4-vinylpyridine)
  diblock copolymer.
\newblock Biotechnol Bioeng 108: 983--987.
\bibAnnoteFile{racine}

\bibitem{thery3}
Azioune A, Storch M, Bornens M, Th\'ery M, Piel M (2009) Simple and rapid
  process for single cell micro-patterning.
\newblock Lab Chip 9: 1640--1642.
\bibAnnoteFile{thery3}

\bibitem{okanoroots}
Okano T, Yamada N, Okuhara M, Sakai H, Sakurai Y (1995) Mechanism of cell
  detachment from temperature-modulated, hydrophilic-hydrophobic polymer
  surfaces.
\newblock Biomaterials 16: 297--303.
\bibAnnoteFile{okanoroots}

\bibitem{werner}
Schmaljohann D, Oswald J, Jorgensen B, Nitschke M, Beyerlein D, et~al. (2003)
  Thermo-responsive pnipaam-g-peg films for controlled cell detachment.
\newblock Biomacromolecules 4: 1733--1739.
\bibAnnoteFile{werner}

\bibitem{okano}
Nagase K, Kobayashi J, Okano T (2009) Temperature-responsive intelligent
  interfaces for biomolecular separation and cell sheet engineering.
\newblock {J} R {S}oc Interface 6: S293--S309.
\bibAnnoteFile{okano}

\bibitem{thick1}
Mizutani A, Kikuchi A, Yamatao M, Kanazawa H, Okano T (2008) Preparation of
  thermoresponsive polymer brush surfaces and their interaction with cells.
\newblock Biomaterials 29: 2073--2081.
\bibAnnoteFile{thick1}

\bibitem{thick2}
Nagase K, Watanabe M, Kikuchi A, Yamato M, Okano T (2011) Thermo-responsive
  polymer brushes as intelligent biointerfaces: Preparation via atrp and
  characterization.
\newblock Macromol Biosci 11: 400--409.
\bibAnnoteFile{thick2}

\bibitem{ito}
Liu H, Ito Y (2002) Cell attachment and detachment on micropattern-immobilized
  poly(n-isopropylacrylamide) with gelatin.
\newblock Lab Chip 2: 175--178.
\bibAnnoteFile{ito}

\bibitem{okano2}
Takahashi H, Nakayama M, Itoga K, Yamato M, Okano T (2011) Micropatterned
  thermoresponsive polymer brush surfaces for fabricating cell sheets with
  well-controlled orientational structures.
\newblock Biomacromolecules 12: 1414--1418.
\bibAnnoteFile{okano2}

\bibitem{okano3}
Yamato M, Konno C, Utsumi M, Okano T (2002) Thermally responsive
  polymer-grafted surfaces facilitate patterned cell seeding and co-culture.
\newblock Biomaterials 23: 561--567.
\bibAnnoteFile{okano3}

\bibitem{okano4}
Williams C, Tsuda Y, Isenberg BC, Yamato M, Shimizu T, et~al. (2009) Aligned
  cell sheets grown on thermo-responsive substrates with microcontact printed
  protein patterns.
\newblock Adv Mater 21: 2161--2164.
\bibAnnoteFile{okano4}

\bibitem{bibi}
Malham IB, Bureau L (2010) Density effects on collapse compression and adhesion
  of thermoresponsive polymer brushes.
\newblock Langmuir 26: 4762--4768.
\bibAnnoteFile{bibi}

\bibitem{Leckband}
Plunkett KN, Zhu X, Moore JS, Leckband DE (2006) Pnipam chain collapse depends
  on the molecular weight and grafting density.
\newblock Langmuir 22: 4259--4266.
\bibAnnoteFile{Leckband}

\bibitem{biggs}
Ishida N, Biggs S (2007) Direct observation of the phase transition for a
  poly(n-isopropylacryamide) layer grafted onto a solid surface by afm and
  qcm-d.
\newblock Langmuir 23: 11083--11088.
\bibAnnoteFile{biggs}

\bibitem{avi}
Halperin A, Kr\"{o}ger M (2011) Collapse of thermoresponsive brushes and the
  tuning of protein adsorption.
\newblock Macromolecules 44: 6986--7005.
\bibAnnoteFile{avi}

\end{thebibliography}

\section*{Figure Legends}

\end{document}